\documentclass[conference]{IEEEtran}
\usepackage{cite}
\usepackage[english]{babel}

\usepackage[T1]{fontenc}

\PassOptionsToPackage{bookmarks=false}{hyperref}
\usepackage{color}
\usepackage{xcolor}
\usepackage{array}
\usepackage{booktabs}
\usepackage{textcomp}
\usepackage{multirow}
\usepackage{amsmath}
\usepackage{url}

\usepackage{pifont}
\usepackage[cm]{sfmath}
 \usepackage{indentfirst}

\usepackage{amsthm}
\usepackage{amssymb}
\usepackage[export]{adjustbox}
\usepackage{verbatim}
\usepackage{graphicx}
\usepackage{mathrsfs} 
\DeclareMathAlphabet{\mathpzc}{OT1}{pzc}{m}{it}
\usepackage{booktabs}

\usepackage{color}

\usepackage[ruled,linesnumbered]{algorithm2e}
\definecolor{seagreen}{rgb}{0.18, 0.55, 0.24}
\SetAlFnt{\small\color{black}\tt}
\LinesNumbered

\def\BibTeX{{\rm B\kern-.05em{\sc i\kern-.025em b}\kern-.08em T\kern-.1667em\lower.7ex\hbox{E}\kern-.125emX}}

\newcounter{defcounter}
\usepackage{caption}
\usepackage[shortlabels]{enumitem}
\usepackage{cite}
\usepackage[ruled]{algorithm2e}
\mathchardef\period=\mathcode`.
\DeclareMathSymbol{.}{\mathord}{letters}{"3B}
\usepackage{tikz}
\tikzstyle{io} = [fill=black,inner sep=2pt,circle]

\graphicspath{ {images/} }

\usetikzlibrary{arrows,positioning}
\usetikzlibrary{shapes.geometric,positioning,shapes,arrows,calc, decorations.pathreplacing, arrows.meta}

\everymath{\displaystyle}
\usetikzlibrary{backgrounds}

\usepackage{pifont}

\makeatletter
\def\endthebibliography{%
	\def\@noitemerr{\@latex@warning{Empty `thebibliography' environment}}%
	\endlist
}

\makeatletter
\newcommand*\bigcdot{\mathpalette\bigcdot@{.5}}
\newcommand*\bigcdot@[2]{\mathbin{\vcenter{\hbox{\scalebox{#2}{$\m@th#1\bullet$}}}}}
\makeatother

\makeatletter

\theoremstyle{plain}

\usepackage[caption=false,font=normalsize,labelfont=sf,textfont=sf]{subfig}
\usepackage{cite} 
\usetikzlibrary{shapes,arrows}
\tikzstyle{line}=[draw] 
\usepackage{algorithmic}
\usepackage{amsmath,varwidth,array,ragged2e}
\usepackage{graphicx}
\usepackage{enumitem}
\usepackage{amsfonts}
\usepackage{mathrsfs}
\usepackage{url}
\usepackage{soul}
\usepackage{xcolor}
\@ifundefined{showcaptionsetup}{}{%
	\PassOptionsToPackage{caption=false}{subfig}}
\usepackage{subfig}
\makeatother
\usepackage{babel}
\providecommand{\theoremname}{Theorem}

\begin{document}

\title{Blockchain-based Firmware Update Scheme Tailored for Autonomous Vehicles}

	\author{%
  Mohamed~Baza\IEEEauthorrefmark{1},
  Mahmoud~Nabil\IEEEauthorrefmark{1},
  Noureddine~Lasla\IEEEauthorrefmark{4},
  Kemal~Fidan\IEEEauthorrefmark{3}, \\
  Mohamed Mahmoud\IEEEauthorrefmark{1},
  Mohamed Abdallah\IEEEauthorrefmark{4}
  \\
  \IEEEauthorblockA{%
    \IEEEauthorrefmark{1}Department of Electrical and Computer Engineering, Tennessee Tech University, Cookeville, TN, USA
  }
  \IEEEauthorblockA{%
    \IEEEauthorrefmark{4}Division of Information and Computing Technology, College of Science and Engineering, HBKU, Doha, Qatar
  }
  \IEEEauthorblockA{%
    \IEEEauthorrefmark{3}Department of Electrical Engineering and Computer Science, University of Tennessee, Knoxville, TN, USA
  }
}

\maketitle
	
 \begin{abstract}
	Recently, Autonomous Vehicles (AVs) have gained extensive attention from both academia and industry.  AVs are a complex system composed of many subsystems, making them a typical target for attackers. Therefore, the firmware of the different subsystems needs to be  updated to the latest version by the manufacturer to fix bugs and introduce new features, e.g., using security patches. In this paper, we propose a distributed firmware update scheme for the AVs' subsystems, leveraging blockchain and smart contract technology. A consortium blockchain made of different AVs manufacturers is used to ensure the authenticity and integrity of firmware updates. Instead of depending on centralized third parties to distribute the new updates, we enable AVs, namely distributors, to participate in the distribution process and we take advantage of their mobility to guarantee high availability and fast delivery of the updates.
	To incentivize AVs to distribute the updates, a reward system is established that maintains a credit reputation for each distributor account in the blockchain. A zero-knowledge proof protocol is used to exchange the update in return for a proof of distribution in a trust-less environment.
    Moreover, we use attribute-based encryption (ABE) scheme to ensure that only authorized AVs will be able to download and use a new update.
	Our analysis indicates that the additional cryptography primitives and exchanged transactions do not affect the operation of the AVs network. Also, our security analysis demonstrates that our scheme is efficient and secure against different attacks.

	\end{abstract}
\section{Introduction}
\subsection{Motivation}
Over the last few years, the automobile industry has achieved a notable leap towards the realization of practical Autonomous Vehicles (AVs). AVs are equipped with sophisticated systems and subsystems to provide vehicles with advanced communication capabilities, computer vision, autonomous decision-making capability, etc., to enable them to autonomously drive without any human intervention \cite{EV110}. AVs have the potential to enhance our current transportation system by reducing congestion and travel time, increasing fuel efficiency, and improving road safety~\cite{eriksson2008cabernet}.

AVs are composed of many subsystems running specific firmware programs that enable performing all control, monitoring, and data manipulation operations. However, by controlling the functionality of the subsystems through the installation of  infected versions of the corresponding firmwares, an attacker can successfully hack AVs and fully/partially access them, e.g., to involve the vehicle in accidents deliberately, which may lead to dramatic damages and kill people. As an example of this attack, Chrysler company announced a recall for 1.4 million vehicles after hackers have managed to turn-off the engine remotely while the vehicles were on motion by exploiting a hackable software vulnerability via the internet-connected entertainment system~\cite{CHRYSLER}. Therefore, ensuring the integrity and authenticity of AVs' firmware update is primordial and must be carefully addressed. In addition, it may happen that multiple AVs with their various subsystems need to be updates urgently and simultaneously, e.g., to fix  newly discovered bugs, thus a high availability of the updates is required.

Most of the existing solutions for firmware update depends on the client-server model in which a manufacturer delegates the process of firmware distribution to trusted cloud providers, such as Microsoft Azure and IBM Cloud~\cite{lin2009cloud}. However, this central client-server architecture has the single point of failure problem. In case the server is not available, the clients (AVs) cannot access the resources (updates) no matter how powerful the server is. For AVs, there are several factors that make the availability and security of the firmware updates challenging tasks. To elaborate, the number of autonomous vehicles on roads is expected to reach 20.8 million in the U.S. alone [4]. Also, each AV has many subsystems that run different programs designed to accomplish specific functions. This creates tremendous load on the server side and can broaden the sources of cyberattacks. Moreover, AVs are designed to last for many years (15 to 20 years); thus the integrity and authenticity of the firmware should be guaranteed throughout the years of service.

Recently, blockchain with its capability to provide a verified, transparent and distributed ledger without a need of a trusted third party, has drawn the attention of both academia and industry across a wide range of domains, including health-care, finance, and energy~\cite{fernandez2018review}. Moreover, blockchain paved the way to build smart contracts, which serve as a piece of code on the blockchain that can perform an action once specific criteria are satisfied. More importantly, self-enforcing smart contracts can be executed without the need for trusted intermediaries~\cite{christidis2016blockchains}.

\subsection{Contribution}
In this paper, we propose a blockchain-based firmware update scheme tailored for AVs. We use blockchain and smart-contract technology to guarantee the authenticity and integrity of new updates. We also exploit the AVs' inter-communication capability and incetivize AVs to particiapte in the distribution and transfer of new firmware updates from one to another therefore ensuring, high availability and fast delivery of the updates. The main contributions of the proposed scheme are outlined as follows:

\begin{itemize}
\item A consortium blockchain created by multiple manufacturer organizations is proposed. Each consortium member has the permission to  write a smart contract that handle the logic ensuring the authenticity and integrity of its firmware updates without the need for a trusted third party. 

\item A high availability and reliability  is ensured by incentivizing AVs to participate in the distribution of the firmware updates. Distributor AVs are rewarded for their honest participation and the smart contract  is used to manages the reward system and keeps track of the reputation credit of each AV.

\item
Attribute-based encryption (ABE) technique is used to allow manufacturers to set a policy about who have the rights to download and use an update. The access policy is defined on the smart contract that enforces its execution without an intermediary, so only authorized  AVs can  request and receive the update. 

\item Since AVs do not mutually trust each other, a Zero-Knowledge Proof protocol is employed. Each distributor can exchange an encrypted version of the update in return for proofs of distribution from receiver AVs. The delivery of the decryption key is guaranteed by the smart contract which will reveal the key once the proofs are collected. The smart contract also increment the distributor's reputation based on the received proofs. 


 
\end{itemize}

The rest of this paper is organized as follows. 
In Section~\ref{Preliminaries}, we discuss some preliminaries. 
Then, our proposed scheme is presented in details in Section~\ref{proposed}. 
Performance evaluations are provided in Section~\ref{performance}.
Section~\ref{related} discusses the related work.
Finally, we give concluding remarks in Section~\ref{conclusion}.

\section{Preliminaries}

\label{Preliminaries}
In this section, we present the necessary background on blockchain, smart contracts and some cryptographic tools that we have used for this research, as well as the notation used along this paper.
\subsection{Blockchain and Smart Contracts}
Blockchain was first introduced in 2008 as the underline technology behind the cryptocurrency known as Bitcoin~\cite{nakamoto2008Bitcoin} to help make peer-to-peer exchange of value without a centralized third  party.
A blockchain is a distributed, immutable, and append-only data structure formed by a sequence of blocks that are chronologically and cryptographically linked togather~\cite{christidis2016blockchains}. 
Fundamentally, a network composed of a set of nodes called miners or validators are responsible of keeping a trustworthy record of all transactions  through a consensus algorithm in a trust-less environment. To exchange some coins from one account to another, for instance, a new transaction is generated and broadcast to the network. Each user is identified by a pseudonym address, usually generated from its public key, and the transaction is authenticated through a digital signatures computed using the user's private key. 
 One of the exciting applications of blockchains is smart contracts, which are defined as computer codes running on top of a blockchain and is correctly executed without fraud or any interference from a third party~\cite{szabo1997idea}. 
  Each contract has a unique address on the blockchain to identify itself and to allow users or other contracts to interact with it. 
  The most popular smart contracts platform is Ethereum~\cite{wood2014ethereum}, and the de-facto language for creating contracts in Ethereum is Solidity\footnote{https://solidity.readthedocs.io/en/develop/}.


\begin{figure*}[ht]
    \centering
       \includegraphics[width=0.99\textwidth]{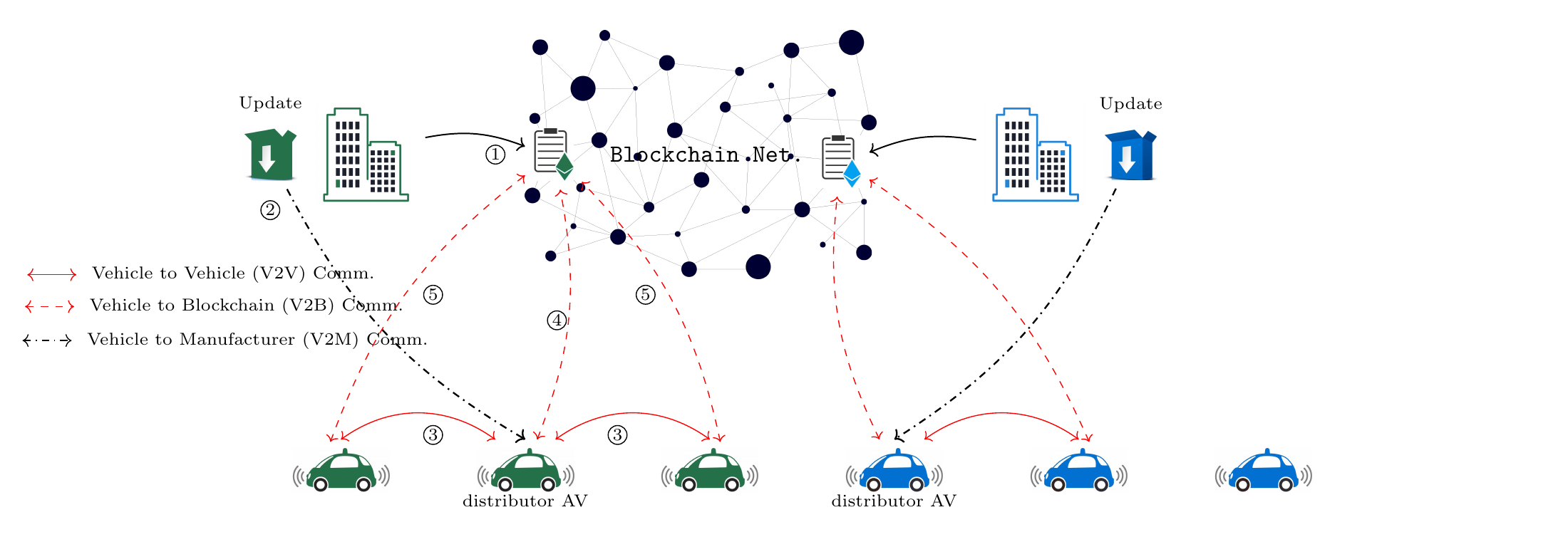}
  \caption{System architecture: (1) The manufacturer creates a smart-contract for a new firmware update by including its hash code for authenticity checking by AVs. (2) The manufacturer sends the new update to top-reputation AVs (distributors). (3) A distributor exchanges an encrypted version of the update in return for proof of reception of the update by a responder AV. (4) A redeem transaction, containing multiple proofs, is sent to the smart contract to update the distributor's reputation. (5) The responder AV receives the decryption key of the firmware update from the smart contract.}
\label{model}
\end{figure*}

\begin{table}[ht]
 \centering
 \caption{System Notations.}
 \begin{tabular}{l|l}
 \hline \hline
  Symbol                                       & Description \\ \hline
  $\mathcal{M}_\theta$                         & A manufacturer company for AVs.\\ \hline
  $PK_\theta/ SK_\theta$                       & Public/ Private key pair for manufacturer $\mathcal{M}_\theta$. \\ \hline
  $PK_{V_{j}}/ SK_{V_{j}}$                     & Public/ Private key pair for vehicle $V_j$. \\ \hline
  $\mathcal{PK}_{U_{i}}/ \mathcal{VK}_{U_{i}}$ & zk-SNARK proving/verifying key pair for update (${U}_i$).\\ \hline
  ${U}_i$                                      & $i$th firmware update version. \\ \hline
  $P_i$                                        & Access policy defined by the manufacturer for $U_i$. \\ \hline
  $AC_i$                                       & Authentication code of update ${U}_i$ and policy (${P}_i$). \\ \hline
  $V_j$                                        & A responder vehicle $j$ that receives an update ${U}_i$.\\ \hline
  $k_j$                                        & Encryption Key for ${U}_i$ of vehicle $V_j$. \\ \hline
  ${h}_i$                                      & Hash of $k_j$.  \\ \hline
  $\hat{U}_i$                                  & The firmware update (${U}_i)$ encrypted with $k_j$.  \\ \hline
  $\mathcal{C}_{V_{j}}$                        & A concatenation of $AC_i$ and $h_j$.\\ \hline
  $\sigma_{j}$                                 & A signature of receiving update ${U}_i$ from vehicle $V_j$.\\ \hline
 \end{tabular}
 \label{not}
\end{table}

\subsection{Cryptographic Tools}
 The notation details  used in the remaining paper  are listed in Table.~\ref{not}.
 
\subsubsection{Attribute Based Encryption}
Attribute based encryption (ABE) is an encryption scheme that allows access control over encrypted data. In ABE, each user is assigned a set of secret keys corresponding to his/her set of attributes. 
Then, a message is encrypted under an access policy formed from the system's set of attributes. The message can only be decrypted by the users who have the attributes that can satisfy the policy. In our scheme, we use the attribute based encryption scheme proposed in \cite{rouselakis2015efficient} to enable the distributor AV to identify the neighbouring AVs who have the required features to download a firmware update. 
This scheme is a ciphertext-policy attribute-based encryption (CP-ABE), where the access policy is embedded in the ciphertext. 

\subsubsection{Zero-Knowledge Succinct Non-Interactive Argument of Knowledge (zk-SNARK):} 
zk-SNARK is a proof construction in which one, called the prover, can prove possession of a specific information, called a witness ($w$), e.g., a secret key, to someone else, called the verifier, without revealing that information. zk-SNARK does not require any interaction between the prover and verifier. Moreover, these schemes are efficient in the sense that the zero-knowledge proof can be verified quickly.

We adopt the zk-SNARK scheme in~\cite{ben2014succinct}. Formally speaking, let $L$ be an NP language with $C$ as its decision circuit. 
Two keys play an essential role, namely, the proving key ($\mathcal{PK}$) and the verifying key ($\mathcal{VK}$). 
The proving key allows any prover to compute a proof $\pi$ for a statement $y \in L$ with a witness $w$. 
Typically, a zk-SNARK scheme consists of the following three polynomial-time algorithms:

\begin{enumerate}
    \item $Gen(1^\lambda,C) \xrightarrow{} (\mathcal{PK}, \mathcal{VK})$. Given a security parameter $\lambda$ and $C$ as a decision circuit, the $Gen$ algorithm generates two public keys, including $\mathcal{PK}$ and $\mathcal{VK}$, that are used to prove/verify the membership in $L$.     
    
    \item $Prove(\mathcal{PK}, y, w) \xrightarrow{} \pi$: Given $\mathcal{PK}$, instance $y$, and witness for a NP statement $w$, 
    the $Prove$ algorithm generates a proof $\pi$ for the statement $x \in L_{c}$.
    
    \item $Verify(\mathcal{VK}, y,\pi ) \xrightarrow{} \{0, 1\}$. Given $\mathcal{VK}$, instance $y$, and the proof $\pi$, 
    the $Verify$ algorithm outputs 1  if $y \in L_{c}$, allowing the verifier to verify the instance $y$. 
\end{enumerate}

\subsubsection{Aggregate Signatures}
 Given $n$ signatures ($\sigma_{1},\dots,\sigma_{n}$) on $n$ distinct messages from $n$ users, aggregate signature scheme can be used 
 to aggregate all these signatures into a single short signature ($\sigma_{agg}$). 
Then, given $\sigma_{agg}$ and the $n$ messages, a verifier can efficiently ascertain that the $n$ users indeed signed the messages. 
In our scheme, we use the aggregate signature scheme proposed in~\cite{boneh2003aggregate} to reduce computations overhead on the blockchain.  
The idea is that instead of sending a transaction to the blockchain each time a distributor AV distributes a firmware update and gets a proof from other AV, it can aggregate several proofs 
to create one short aggregated proof to reduce the number of transactions sent to the blockchain.

\begin{figure*}[ht]

    \centering
 
       \includegraphics[width=0.99\textwidth]{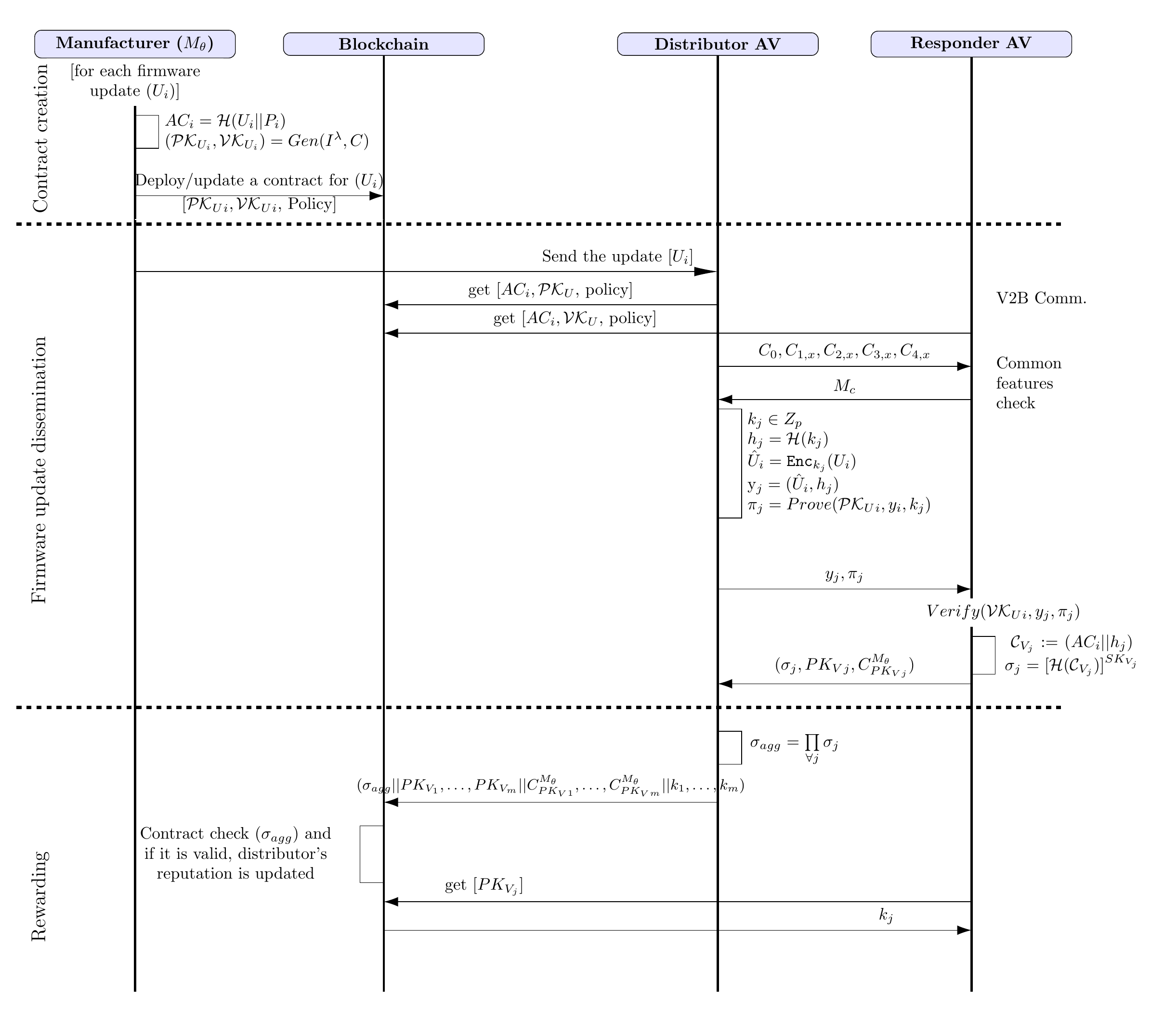}
    \caption{Firmware update scheme sketch.}
    \label{sketch}
\end{figure*}

\section{Secure and Scalable Firmware Update Scheme} \label{proposed}
In  this  section,  we  present our scheme that aims to ensure secure and scalable delivery of firmware updates from automobile manufacturers to AVs. 
We first present a general architecture for the system, followed by system initialization, the smart contract creation, firmware update dissemination, and rewarding. 

\subsection {System Architecture}
Fig.~\ref{model} shows the system architecture, which is comprised of two manufacturers with their AVs, two smart contracts for firmware updates of each manufacturer, and a consortium blockchain. A sketch of the possible interactions between the different system entities is shown in Fig.~\ref{sketch}.
The role of each entity in the system is discussed in the following paragraphs.

\vspace{6pt}
\noindent \textit{Manufacturer.} 
The manufacturer is responsible for keeping its manufactured AVs updated with the latest versions of the different firmware updates for the subsystems that control the AVs.  During the manufacturing of AVs, the manufacturer uploads each AV with a set of cryptographic keys and public parameters that will be used to ensure secure distribution of firmware updates. Also, each time a new update is released, a corresponding smart contract is deployed by the manufacturer to allow AVs to check the integrity and authenticity of the update. In addition, to attract AVs to participate in the distribution of an update, the manufacturer compensates the participants through a rewarding mechanism, e.g., momentary rewards and free or reduced-price maintenance services. 

\vspace{6pt}
\noindent \textit{Autonomous Vehicles.} 
We distinguish between two types of AVs, \textit{distributors} and \textit{responders}. 
The distributor AV disseminates a new firmware update to other AVs (responders) in its vicinity. 
Each responder AV that receives an update can also act as a distributor of that update.
By this way, we can ensure the large-scale dissemination of the update quickly. 
Initial distributors are selected by the manufacturer based on their reputations which are recorded in a smart-contract.

\vspace{6pt}
\noindent \textit{Smart Contract.} 
For each new firmware update, a smart contract is created. 
The contract contains the necessary credentials allowing any receiver of the update to authenticate it and verify its integrity. 
In addition, the contract implements the reputation logic that evaluates and keeps track of the AVs' activities in the distribution of the firmware update. 
More specifically, the contract increases the reputation of a distributor AV after receiving proofs of the firmware distributions from responder AVs. 

\vspace{6pt}
\noindent \textit{Blockchain Network.}  
Blockchain network is at the center of our system and it executes the smart contracts in a distributed manner without relying on a central party. 
This is mandatory to ensure a scalable and secure firmware update dissemination. 
Moreover, to improve the efficiency of the system, we opt for a \textit{consortium blockchain}, where the validators, i.e., nodes with write permission on the shared ledger, are known and trusted entities. 
In our case, the validators can be the manufacturers of the different automobile brands. 

\subsection{System Initialization}
\label{System Initialization}
A multi-authority attribute based encryption scheme is used, where each manufacturer is considered as an authority that decides a set of attributes (or features) for its AVs.  
$\mathbb{M}$ is the set of all available manufacturers and a manufacturer $\mathcal{M}_\theta \in \mathbb{M}$. 
Let $\mathbb{A}$ be the set of all attributes (or features) in the system, an access policy ($A, \delta $) on $\mathbb{A}$ with $A$ $\in \mathbb{Z}_{p}^{l \times n} $, called the  share generating matrix in the field $\mathbb{Z}_{p}$ of prime order $p$ with $ l $ rows and $ n $ columns, and a function $ \delta $ that labels the rows of $A$ with attributes from $\mathbb{A}$, i.e., $ \delta: [l] \rightarrow \mathbb{A} $. 
In addition, let $ \rho $ be a function that maps attributes in rows to its manufacturers, where $\rho: [l] \rightarrow \mathcal{M}_\theta $.
 Consider $e: \mathbb{G}_1 \times \mathbb{G}_2 \rightarrow G_T$ a cryptographic bilinear map with generators $g_1 \in \mathbb{G}_1$ and $ g_2 \in \mathbb{G}_2$, where $\mathbb{G}_1$ and $\mathbb{G}_2$ are multiplicative group. 
 Each manufacturer $\mathcal{M}_\theta \in \mathbb{M}$  should select two random elements $(\alpha_{\theta}, y_{\theta}) $ $ \stackrel{R}{\leftarrow} $  $ \mathbb{Z}_{p}^{*}$ as its secret keys, and then, $\mathcal{M}_\theta$ can compute  its public key as \textit{PK$_\theta$} = $\{e(g_1, g_2)^{\alpha_{\theta}}, g_1^{y_{\theta}}\} $. 
 Besides, a public hash function ${\mathcal{H}:\{0,1\}^*\rightarrow G_1}$ is used to map an AV global identifier $GID$ to a point in $G_1$, public hash function ${F:\{0,1\}^*\rightarrow G_1}$ that maps an attribute $a \in \mathbb{A}$ to $G_1$, and a function $T$ that maps an attribute $a \in \mathbb{A}$  to the manufacturer $\mathcal{M}_\theta$, hence, the function $ \rho (\cdot)$ can be redefined as $ \rho (\cdot)$: T($ \delta (\cdot) $) . The global  parameters are then defined as $GP=\{ \mathbb{G}_1,\mathbb{G}_2,G_T,\mathbb{Z}_p, \mathcal{H},F,T,\mathbb{A},\mathbb{M}\}$.
 
Besides, during the production, $\mathcal{M}_\theta$ should assign each AV a key for each assigned attribute $a \in \mathbb{A}$ using the AV global identity $GID$ as follows: $\mathcal{M}_\theta$ chooses a random $ t  \stackrel{R}{\leftarrow} $ $,\mathbb{Z}_{p}^{*}$ and outputs to the AV attributes secret keys as $SK_{GID, a} = \{{K_{GID, a} = g_2^{\alpha_{\theta}}H(GID)^{y_{\theta}}F(a)^{t}}, K'_{GID, a} = g_1^{t}\}$. Finally, for each AV, $\mathcal{M}_\theta$ generates a public/private key pair as follows: a random number $x_a \stackrel{R}{\leftarrow} \mathbb{Z}_{P}$ is selected as the private key and the corresponding public key is $PK_{V_{j}}=g_{2}^{x_a}$. The public key $PK_{V_{j}}$ should be associated with a manufacturer's certificate as ($C^{M_\theta}_{PK_{V_j}})$. Then, the public/private key pair ($PK_{V_{j}}, x_a$)  and $C^{M_\theta}_{PK_{V_j}}$ should be added to the AV's tamper proof device along with manufacturer's public key $PK_\theta$.

\begin{algorithm}[!t]
\SetKwProg{Fn}{function}{}{}
\SetKwProg{Contract}{contract}{}{}
\SetKwData{NumOfUpdatedObjects}{numOfUpdatedObjects}
\SetKwIF{If}{ElseIf}{Else}{if}{}{else if}{else}{end if}
\SetKwFunction{FirmwareUpdateContract}{FirmwareUpdateContract}
\SetKwFunction{FirmwareUpdate}{FirmwareUpdate}
\SetKwFunction{RecieveProof}{RecieveProof}
\SetKwFunction{UpdateReputation}{UpdateReputation}

\Contract{\FirmwareUpdate}{
\textcolor{blue}{mapping}{(\textcolor{blue}{address} => int) Reputation}
\tcp{Mapping for distributors reputation}
\textcolor{blue}{mapping}{(\textcolor{blue}{address} => int) UpdatedAVs}
 \tcp{Mapping for AVs with the No.of obtained updates}
  \BlankLine
  \Fn{\FirmwareUpdate{\_PK, \_VK, \_AC\textsubscript{i},  \_P\textsubscript{i}, X}}{
    
        PK $\leftarrow$ \_PK \tcp{Proving Key} 
        VK $\leftarrow$ \_VK \tcp{Verifying key}
        AC\textsubscript{i} $\leftarrow$ \_AC\textsubscript{i} \tcp{authentication code} 
        P\textsubscript{i} $\leftarrow$ \_P\textsubscript{i} \tcp{ABE Policy}
        MaxUpdate $\leftarrow$ X \tcp{Max. No. of download per Update}

  }
  \BlankLine

  \Fn{\RecieveProof{$\sigma_{agg}$, \texttt{PK}[$\>$], \texttt{C}[$\>$]  \texttt{keys}[$\>$]}}{

   \textcolor{blue}{address} [] RecievedAVs \tcp{Received AV list}
    \For{$s\leftarrow 0$ \KwTo PK.lengh}{ 
    \If {verifySig(pk\_M, PK[s], C[s])} {return} 
     \If {UpdatedAVs[PK[s]] > MaxUpdate} {return}

     $h_s\leftarrow \mathcal{H}(\texttt{keys}[s]) $ \\
     $\mathcal{C}_{V_{s}}\leftarrow \mathcal{H}(AC_i,h_s)$
     RecievedAVs.push(Pairing$(PK[s], \mathcal{C}_{V_s})$))

    }
    \If{Pairing$(g_{1},\sigma_{agg})$=Prod(RecievedAVs)}{
     
      UpdateReputation(msg.sender, PK.length)

        \For{$i\leftarrow 0$ \KwTo PK.lengh}{ 
     emitEvent("KeyRevealed", PK\_{i}, \texttt{keys}[i]) 
  
     UpdatedAVs[PK\_{i}]$\leftarrow $ UpdatedAVs[PK\_{i}]+1
     
  } } }
     
  \BlankLine
  \Fn{\UpdateReputation{Dist, N}}{
  \tcp{increase reputation  distributors}
      Reputation[Dist]$\leftarrow $ Reputation[Dist]+=N
  }
}
\caption{Pseudocode for the \textit{Firmware Update} contract}
\label{alg:contract}
\label{alg1}
\end{algorithm}
\DecMargin{1em}

\subsection{Smart Contract Creation}
Upon releasing a new update by a subsystem's manufacturer, denoted by $U_{i}$, the AV manufacturer that uses the subsystem should first test the update. Note that a subsystem's manufacturer may be different from the AV manufacturer. If the AV manufacturer decides to use it on its AVs, it starts the firmware update as follows. The manufacturer creates a smart contract and initializes it by two attributes:
(1) A proving/verifying key pair: $\mathcal{(PK}_{U_{i}}, VK_{U_{i}})= Gen(1^{\lambda}, C)$, required for the execution of the zk-SNARK protocol; (2) An authentication code for the new firmware update: $AC_i=\mathcal{H}(U_i || P_i)$, where $P_i$ is the access policy defined by the manufacturer to deliver the update to only AVs that have the features defined in the policy.

\noindent The manufacturer deploys a smart-contract by broadcasting a transaction to the blockchain network. The deployed smart-contract is described in Algorithm \ref{alg1} and includes the following main functions:

\begin{itemize}
    \item \textit{Authenticity and integrity of a firmware update.} Since the update's authentication code and verification key are stored in the contract, AVs, by consulting the blockchain, can check whether a received firmware update is the same one that was originally approved by the AV manufacturer. 
    
    \item \textit{AVs' reputation}. When an AV participates in the distribution of a new update, the proof of distribution is sent to the smart contract which in turn increases its reputation. The manufacturer rewards the highly-reputed AVs, i.e., the active AVs in distributing the firmware. The reward can be momentary, free or reduced-price maintenance service, etc.
    
    \item \textit{Firmware access control}. 
    Each firmware update has an access policy set by the manufacturer, and the AVs that have enough features to satisfy the policy can receive the firmware. This can restrict the distribution of the firmware to only certain AVs. The access policy of an update is included in the update's smart-contract. 
\end{itemize}

\subsection{Firmware Update Dissemination}
In this stage, the manufacturer starts the dissemination process of a new update by first selecting the most active AVs in distributing updates (based on their reputations) to act as the initial distributors. 
As discussed before, rewards are used to incentivize the AVs to act as distributors and actively distribute the new firmware, 
but the rewarding system should be secure to ensure that only honest distributors which distributes the firmware are rewarded. 
In practice, each AV manufacturer will use many subsystems made by different companies, and therefore, it is very frequent that different AV manufacturer may use the same subsystems produced by the same company in their vehicles. 
Thus, it is very important for each manufacturer to ensure that its distributors will deliver a particular update to only certain models of its AVs. 


Thanks to ABE, as presented in Section \ref{Preliminaries}, each manufacturer can define an access  policy for each update on the associated smart contract, where only AVs that belong to the same manufacturer and have enough fractures, such as model, year of manufacturer, etc, can decrypt and use the firmware update they got from a distributor.  

For a distributor to find other AVs which can satisfy the access policy of an update and deliver it, the following steps should be taken. 
\begin{enumerate}
    \item A distributor AV first queries the blockchain for the $AC_i$, proving key $\mathcal{PK}_{U_{i}}$, and the manufacture's access policy ($A,\delta$).
    \item Then, distributor AV should broadcast an encrypted challenge message ($M_c$) using the ABE to the nearby AVs. This message is encrypted using the manufacture's public key $PK_\theta$ under the access policy ($A,\delta$) set by the manufacturer. 
    Hence, only the AVs which owns the set of attributes that satisfy the policy are able to decrypt the ciphertext $CT$. 
    To encrypt $M_c$, distributor AV  first creates two random vectors $v = ({z, v_{2}, \cdots v_{n}})^{T}$ and $w = ({0, w_{2}, \cdots w_{n}})^{T}$, where \{${z, v_{2}, \cdots v_{n}, w_{2}, \cdots w_{n}}$\} are elements randomly selected from $Z_{p}^{*}$. 
    We denote $\lambda_{x}$ as the share of the random secret $z$ corresponding to row $ x $, i.e., $\lambda_{x} = (A_{x} \cdot v)$ and $w_{x}$ denotes the share of zero, i.e., $w_{x} = (A_{x} \cdot w)$, where $A_{x}$ is the x-th row of access matrix $A$. The distributor AV chooses a random element $t_{x}$ $ \stackrel{R}{\leftarrow} $ $\mathbb{Z}^*_{p}$ for each row in the policy matrix $A$ and computes the $CT$ as:
    \begin{figure}[!h]

    \centering
 
       \includegraphics[width=0.3\textwidth]{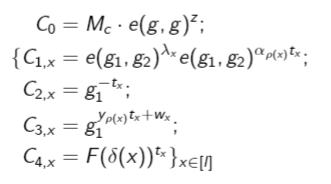}
   
\end{figure}

    \item After a responder AV receives $CT$, it first queries the smart contract for the access policy $(A, \delta)$, $\mathcal{VK}_{U_i}$ and $AC_i$. Then, to decrypt $M_c$, the AV should use the policy $(A, \delta)$ from the blockchain and its secret keys $(K_{GID, a}, K'_{GID, a})$ for the subset of rows $A_{x}$ of satisfied attributes and for each row $ x $ to compute

     \begin{figure}[!h]

    \centering
 
       \includegraphics[width=0.43\textwidth]{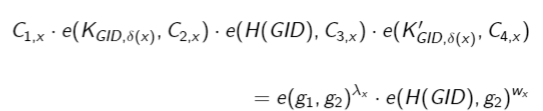}
   
\end{figure}

    Then, AV calculates the constants $c_{x} \in \mathbb{Z}^*_{p}$ such that $ \Sigma_{x} c_{x} A_{x} $ = (1, 0, . . . 0) and computes:
    \begin{center}
        $\Pi_{x}(e(g, g)^{\lambda_{x}}. e(H(GID), g)^{w_{x}})^{c_{x}} = e(g, g)^{z} $
    \end{center}
    This is true because $\lambda_{x} = (A_{x} \cdot v)$ and $w_{x} = (A_{x} \cdot w)$, where $ \langle(1, 0, \cdots , 0) \cdot v \rangle = z $ and $ \langle(1, 0, \cdots , 0) . w \rangle = 0 $.
    Hence, the challenge message can be decrypted as $M_c = C_{0}/e(g, g)^{z}$.
    \item Finally, once a responder AV manages to get $M_c$, it then replies to the distributor AV with the correct $M_c$. Henceforth, both the distributor and responder AV can proceed with the firmware update transfer.
\end{enumerate}


A distributor AV sends the firmware update to a responder AV in return for a signature, i.e., a proof for disseminating the firmware. 
This exchange of firmware update and proof can be made in a trust-less way using zk-SNARK protocol as follows.
\begin{enumerate}
\item The distributor AV generates a secret key $k_j \in \mathbb{Z_{P}}$ and calculates $h_{j}=\mathcal{H}(k_{j})$.
\item Then, it computes  $\hat{U}_i=\mathtt{Enc}_{k_{j}}(U_i)$, where $\mathtt{Enc}$ is a symmetric-key encryption algorithm.
\item For zk-SNARK protocol, the secret witness is the instance $y_{j}= (\hat{U}_i, h_j$) and $k_j$. The NP statement is as follows: 
\begin{equation}
\label{NP}
\exists\> k_{j} : \mathcal{H}(k_{j})=h_{j}\>\land\> \mathcal{H}(Dec_{k_{j}}(\hat{U_{i}}), P_i)=AC_{i}
\end{equation}

 Which attests that the distributor AV has a key, $k_{j}$, such that its hash is $h_{j}$, and if $k_{j}$ is used to decrypt $\hat{U_{i}}$, it will match the update authentication code $AC_{i}$. After that, the distributor computes a zero-knowledge proof $\pi_{j}= Prove(\mathcal{PK}_{U_i}, y_{j}, h_{j})$ and sends $ y_{j} ||\pi_{j}$ to the responder.

 \item Upon receiving ($y_{j} ||\pi_{j}$), the responder first verifies that $Verify(\mathcal{VK}_{U_i}, y_{j}, \pi_{j})=1$, and then computes a signature $\sigma_{j}=[\mathcal{H}(\mathcal{C}_{V_{j}})]^{SK_{V_{j}}}$, where $SK_{V_{j}}$ is its private key  and $\mathcal{C}_{V_{k}}=(AC_i,h_j)$.
\item Finally, it sends ($\sigma_{j} || PK_{V_{j}} || C^{M_\theta}_{PK_{V_j}}$)  to the distributor.

\end{enumerate}

\subsection{Rewarding}
In this phase, the distributor AV sends a redeem transaction, containing multiple proofs, to the smart contact to update its reputation proportionally to the number of AVs which received the update. This rewarding process is done as follows.
\begin{enumerate}
\item  To reduce the number of transactions sent to the blockchain, instead of making a transaction each time a firmware is transferred, one transaction can be sent for several firmware transfers efficiently, as follows. Once the distributor AV gets the proofs of transferring an update ($\sigma_{j}$) from other vehicles, it aggregates multiple signatures into a single signature ($\sigma_{agg}$) as follows: $\sigma_{agg}= \prod\limits^{}_{\forall{j}} \sigma_{j}$. Note that, a signature $\sigma_{j}$ should be different from other received signatures. In other words, the distributor should generate a distinct $k_j$ for each time he sends the new update to other vehicles. 
 \item The distributor sends a transaction to the blockchain containing ($\sigma_{agg} || PK_{V_{1}},\ldots,PK_{V_{m}}||C^{M_\theta}_{PK_{V_1}},\ldots, C^{M_\theta}_{PK_{V_m}}|| k_{1},\ldots, k_{m})$ where $m$ is the number of vehicles that received the update.
 \item The smart contract method \texttt{RecieveProof} first verifies that the received public key is one of the certified keys by the manufacturer (see \texttt{verifysign} in Algorithm~\ref{alg:contract}) as well as many number of times that AV gets the update. Then, it computes $h_j=\mathcal{H}(k_j)$ and $\mathcal{C}_{V_{j}}=\mathcal{H}(AC_i,h_j)$ for all $j$.
Thereafter, it verifies the aggregated signature by checking if $e(g_1,\sigma_{agg}) = \prod\limits^{}_{\forall{j\in m}}e(PK_{V_{j}}, \mathcal{C}_{V_{j}})$ or not (see \texttt{pairing} in Algorithm~\ref{alg:contract} which can be executed by a pre-compiled contract for elliptic curve pairing operations available at ~\cite{boneh}). 
\item Finally, the distributor is rewarded by increasing its reputation index proportionally to the number of vehicles that received the update (see the method \texttt{UpdateReputation} in Algorithm~\ref{alg:contract}). 
\item A responder AV queries the contract for a relevant event associated with its public key for the decryption key $k_j$ it needs to decrypt $\hat{U}_i$ to get $U_i$ (see event \texttt{"KeyRevealed"} in Algorithm~\ref{alg:contract}).
\end{enumerate}
\section{Performance Analysis}
\label{performance}
\subsection{Performance Evaluation}
In this section, we evaluate the computation overhead for the cryptography operations used in our scheme.

The computation times of ABE are measured using Intel Core i7- 4765T 2.00 GHz and 8GB RAM machine and Python charm cryptographic library in ~\cite{nabil2018toward}. 
In our scheme, a distributor AV needs to broadcast a challenge packet ($M_c$) encrypted by a number of attributes ($\gamma$) specified in the smart contract. According to~\cite{nabil2018toward}, 
the required time for encryption is (10.9$\times \gamma$ + 1.35) ms. 
In addition, a responder AV needs to decrypt ($M_c$) with total decryption time that can be formulated as (4.03$\times \gamma$ + 0.01) ms. 
After running the ABE scheme, zk-SNARK protocol should be run. 
In this protocol, a proof is generated by the distributor and then verified by the responder. We implemented the NP statement in Equation \ref{NP} using Zokrates\footnote{https://github.com/Zokrates/ZoKrates} toolbox. MIMC \cite{albrecht2016mimc} is used for encryption/decryption due to its efficiency with zk-SNARK proofs, and sha256 is used for hashing. The time to generate the proof is 6 seconds, where as, the verification is 5 milliseconds. It should be noted that the distributor AV can generate multiple the proofs offline before starting the communication session with the responder AVs. Hence, the total computation time needed to run ABE and zk-SNARK verification of our scheme is low, which is suitable for our application because the AVs are in motion and their communication time is short.    
 
In our scheme, blockchain is required to ensure the authenticity and integrity of the new update.
To reduce the cost needed to execute our scheme on the blockchain, most of the computations to secure the scheme are done outside the blockchain. Using a consortium blockchain will remove any constraint on the amount of data that should be sent and stored on the blockchain. 
Additionally, our scheme reduces the on-chain operations by reducing the number of transactions sent to the blockchain by aggregating several firmware transfers in one transaction using aggregate signature scheme. 
Also, as discussed before, the computation cost to run our scheme is low, and according to~\cite{eriksson2008cabernet}, the mean throughput for delivering data to and from moving vehicles that use IEEE 802.11 protocol is equal to 760 kbit/s. 
If we assume that the size of a firmware update equals to 1 MByte, then the time required to transfer the update is 1.3 seconds.
Therefore, given this transfer time and the short time needed for the cryptographic computations, our scheme can be executed during the contact time of two moving AVs.

\subsection{Security analysis}
In this section, we discuss the possible security threats and how our scheme mitigate to each of them.
\begin{itemize}
\item \textit{Firmware integrity.} 
Since the authentication code of each new firmware update is recorded in the smart contract, our scheme resists any attempt by an adversary to distribute malicious updates.
    
\item \textit{Firmware distribution and access control.} To prevent unauthorized AVs from accessing a new firmware update, our scheme allows each AV manufacturer to control the access by defining the list of authorized AVs.
Through ABE access policy, which is registered in the blockchain, a distributor can prescribe the AVs that have the right to receive the update and prevent unauthorized AVs from receiving it. In addition, because the access policy is embedded in the update authentication code ($AC$), any responder AVs (receiver) can decide if it is concerned by a particular update or not. 
If a compromised/malicious distributor AV changes the access policy used to encrypt the challenge message, the responder AV will detect this change during the zk-SNARK verification. 
Hence, the distributor (attacker) will not be able to get a proof of distribution from the responder. 

\item \textit{DoS attack resistance.} The proposed scheme resists Denial-of-service (DoS) attacks~\cite{dolev1983security} that aim to disable the system and the rewarding mechanism. 
This attack is not possible since there is no central unit that distributes the firmware or runs the scheme. For this attack to succeed, the attackers need to control the majority of the validators (manufacturers) of the blockchain network, which is presumably impossible.


\item \textit{Update audibility.} 
In our scheme, the AVs that have distributed or received an update are recorded in the blockchain. This gives the manufacturer an accurate insight about the firmware state of its AVs, i.e., which AVs have received   a particular update.

\end{itemize}

\section{Related Work}
\label{related}


In the literature, the security of firmware update has been discussed in several contexts, including wireless sensor network~\cite{kim2017adaptive,dutta2006securing}, IoT~\cite{lee2017blockchain,leiba2018incentivized}, vehicular network~\cite{nilsson2008secure}, etc.  The existing works can be classified 
either as centralized (client-server model) or decentralized. In the following we review some of the existing solutions in both classes.

In~\cite{nilsson2008secure}, Nilsson et al. proposed a firmware update protocol for modern intelligent vehicles over the internet. The authors suggested a client-server method using a web portal that delivers the firmware update in fragments. The fragments are protected using hash chain and ensuring, therefore, the integrity of updates. However, the system  is  vulnerable to a DoS attack as it relies on a central server.  Also, the central solution does not ensure the availability of the update when several vehicles on the road request the firmware updates at the same time.

In sensor networks, several schemes such as ~\cite{kim2017adaptive},~\cite{dutta2006securing} have been proposed to improve the reliability of delivering new updates/security patches by ensuring their integrity. However, these schemes depend on a single entity to manage the distribution of firmware updates and not scale for large networks. 

In~\cite{lee2017blockchain}, the authors proposed a decentralized solution based on a permission-less blockchain to ensure the integrity of updates by having multiple verification nodes instead of depending on a private centralized vendor network. For the distribution
of updates, a peer-to-peer file sharing network such as BitTorrent is proposed to ensure integrity and versions tractability of updates. 
However, the scheme does not provide any incentive for devices to participate and distribute firmware updates to others.

In~\cite{leiba2018incentivized}, the authors proposed software update framework for Internet of Things (IoT) devices. The framework allows other parties to deliver the updates in return for digital currency paid by the vendor. However, the scheme incurs high financial cost since it depends on Etherum blockchain~\cite{wood2014ethereum} which apply fees for each transaction sent to the network.  
\section{conclusion}
\label{conclusion}
In this paper, a firmware update scheme based on blockchain and smart contract is introduced for autonomous vehicles. A smart contract is used to ensure the authenticity and integrity of firmware updates, and more importantly to manage the reputation values of AVs that transfer the new updates to other AVs. We also use ABE to allow AV manufacturer to target a specific set of AVs that have certain features defined by the manufacturer to download the firmware. A zero-knowledge proof protocol is used to enable the AVs to exchange an update for proof of distribution in a trust-less way. To improve the efficiency, an aggregate signature scheme is used to allow a distributor to combine multiple proofs to make only one transaction on the blockchain when it redeems the rewards. Finally, the smart contract rewards the distributors by increasing their reputation in the blockchain. Our evaluation analysis indicates that the cryptography primitives used to secure the firmware update exchange is suitable to the AVs network. For the future work, we will implement a prototype of our proposed scheme.

\bibliographystyle{IEEEtran}
\bibliography{CC}
	
\end{document}